\documentclass{article}
\usepackage{graphicx} % Required for inserting images

\usepackage[sfdefault]{roboto}
\usepackage[T1]{fontenc}
\usepackage[letterspace=40]{microtype} %Roboto is very dense, this helps spread it out a little

\usepackage{url} %Fixes bibliography url issues
\usepackage{hyperref}

\usepackage{xcolor}
\hypersetup{
    colorlinks,
    linkcolor={red!70!black},
    citecolor={blue!80!black},
    urlcolor={blue!90!black}
}

\usepackage[utf8]{inputenc} %A box package
\usepackage[round]{natbib}
\setcitestyle{open={(},close={)}}
\usepackage[margin=2.5cm]{geometry} 

\usepackage{array} %For table formatting
\usepackage{pifont} %For check and cross marks

\providecommand{\keywords}[1]
{
  \small	
  \textbf{\textit{Keywords---}} #1
}

\usepackage{setspace}
\usepackage[none]{hyphenat} %To remove hyphenation

\begin{document}

\title{From trees to traits: A review of advances in PhyloG2P methods and future directions}
\author{
Arlie R. Macdonald$^{1,2,3,*}$, Maddie E. James$^{2,3}$, Jonathan D. Mitchell$^{1,2}$, Barbara R. Holland$^{1,2}$ \\[2ex]
\small $^1$School of Natural Sciences, University of Tasmania, Hobart, Tasmania, Australia \\
\small $^2$Australian Research Council Centre of Excellence for Plant Success in Nature and Agriculture, \\
\small University of Tasmania, Sandy Bay, Tasmania, Australia \\
\small $^3$School of the Environment, The University of Queensland, Brisbane, Queensland, Australia \\[1ex]
\small $^*$Corresponding author. Email: \href{mailto:arlie.macdonald@utas.edu.au}{arlie.macdonald@utas.edu.au}
}
\maketitle

\begin{abstract}
Mapping genotypes to phenotypes (G2P) is a fundamental goal in biology. So called PhyloG2P methods are a relatively new set of tools that leverage replicated evolution in phylogenetically independent lineages to identify genomic regions associated with traits of interest. Here, we review recent developments in PhyloG2P methods, focusing on three key areas: methods based on replicated amino acid substitutions, methods detecting changes in evolutionary rates, and methods analysing gene duplication and loss. We discuss how the definition and measurement of traits impacts the utility of these methods, arguing that focusing on simple rather than compound traits will lead to more meaningful genotype-phenotype associations. We advocate for the use of methods that work with continuous traits directly rather than collapsing them to binary representations. We examine the strengths and limitations of different approaches to modeling genetic replication, highlighting the importance of explicit modeling of evolutionary processes. Finally, we outline promising future directions, including the integration of population-level variation, as well as epigenetic and environmental information. No one method is likely to identify all genomic regions of interest, so we encourage users to apply multiple methods that are capable of detecting a wide range of associations. The overall aim of this review is to provide practitioners a roadmap for understanding and applying PhyloG2P methods.
\end{abstract}


\lsstyle %Enable microtype letter spacing
\section*{Significance Statement}

PhyloG2P methods have developed rapidly over recent years, and they provide an exciting opportunity to expand genotype to phenotype mapping to trait variation between species. However, PhyloG2P studies have often relied on a simplified understanding of replicated evolution, and some of the common challenges they face have not been synthesised into a clear direction for future research. In this review, we address both of these points by providing a detailed overview of the nature of replicated phenotypic and genotypic evolution, as well as discussing the outstanding challenges PhyloG2P research faces.

\section{Introduction}

Understanding how the genomes of organisms shape their phenotypes is one of the core pursuits of biology. In recent years there have been a growing number of methods that approach this problem from a phylogenetic perspective. This research field is called Phylogenetic Genotype to Phenotype (PhyloG2P) \citep{Smith2020}. PhyloG2P uses evolutionary relationships, as represented by phylogenetic trees, to link changes in genotype with changes in phenotype. Approaching this problem from an evolutionary perspective enables the mapping of genotype to phenotype in situations that would not be possible with typical population genetics approaches \citep{Nagy2020, Smith2020}.

The statistical power of PhyloG2P methods comes from replicated evolution. Replicated evolution (Box 1) is the phenomenon whereby distinct lineages independently evolve similar phenotypes in response to a common environmental pressure. Lineages that have undergone replicated evolution provide natural experiments to uncover the genetic underpinnings of phenotypes. Phylogenetic replication, i.e. the evolution of a phenotype on independent lineages, allows for repeated changes in genotype associated with the phenotype to be distinguished from other lineage-specific genetic changes unrelated to the phenotype of interest. Identifying these correlations between genotype and phenotype has the potential to enhance our understanding of the role that particular genetic elements play in underlying traits and facilitating adaptation \citep{Hiller2012, Chikina2016, Smith2020, Stern2013, Nagy2020}. Naturally, the success of PhyloG2P methods hinges on the assumption that a replicated change in a trait has been caused by similar or identical changes in the genomes of organisms. 

In this review, we begin by covering the aspects of replicated evolution that are essential for understanding PhyloG2P analysis. \citet{Grossnickle2024} and \citet{Thomas2024} identified that PhyloG2P methods would benefit from a more comprehensive consideration of the different biological processes that can cause replication – to this end we first provide a discussion of phenotypic replication and the genomic processes that drive it. We describe existing PhyloG2P methods that focus on three broad categories: single nucleotide polymorphisms (SNPs), edge lengths of phylogenetic trees, and copy number variation. We build on the earlier reviews of \citet{Nagy2020, Smith2020, Pereira2023} and focus on more recent developments in these areas. We conclude the review by identifying key challenges and areas where current analysis could be improved and highlight promising future directions for PhyloG2P development.

\vspace{\baselineskip}

\fbox{\begin{minipage}{0.9\textwidth}

\large \textbf{Box 1: The Language of Replication}

\normalsize The independent evolution of similar traits or phenotypes in response to similar environmental pressures is commonly known as parallel or convergent evolution, depending on the similarity of the ancestral species or lineages. On a phylogenetic scale, parallel evolution typically refers to independent evolution of similar phenotypes in ``closely'' related species, while convergent evolution occurs in more ``distantly'' related species. At the genetic sequence level, parallel evolution is often used to describe identical nucleotide substitutions from a shared ancestral state, while convergent evolution involves different ancestral nucleotide states. At the genetic network scale, parallel evolution involves changes in homologous genetic elements, whereas convergent evolution involves changes in distinct elements of the network (for more details, see \citet{Pereira2023}). However, it is not always clear where the boundaries between these categories of close/similar and distant/distinct lie. Additionally, some systems may contain both parallelism and convergence at different levels, such as distantly related species (convergent on the phylogenetic level) which both have had identical substitutions at a particular nucleotide position from the same ancestral state (parallel on the genetic sequence level). We wish to avoid potential confusion caused by the language of parallelism and convergence. Following \citet{James2023}, we use the term ``replicated evolution'' to refer to all forms of independent evolution of similar phenotypes in response to shared environmental pressures, regardless of the underlying genetic mechanisms.

\end{minipage}}

\subsection{Defining Traits and Phenotypes} \label{subsec:phenotypic_replication}

\begin{figure}
    \centering
    \includegraphics[width=1\linewidth]{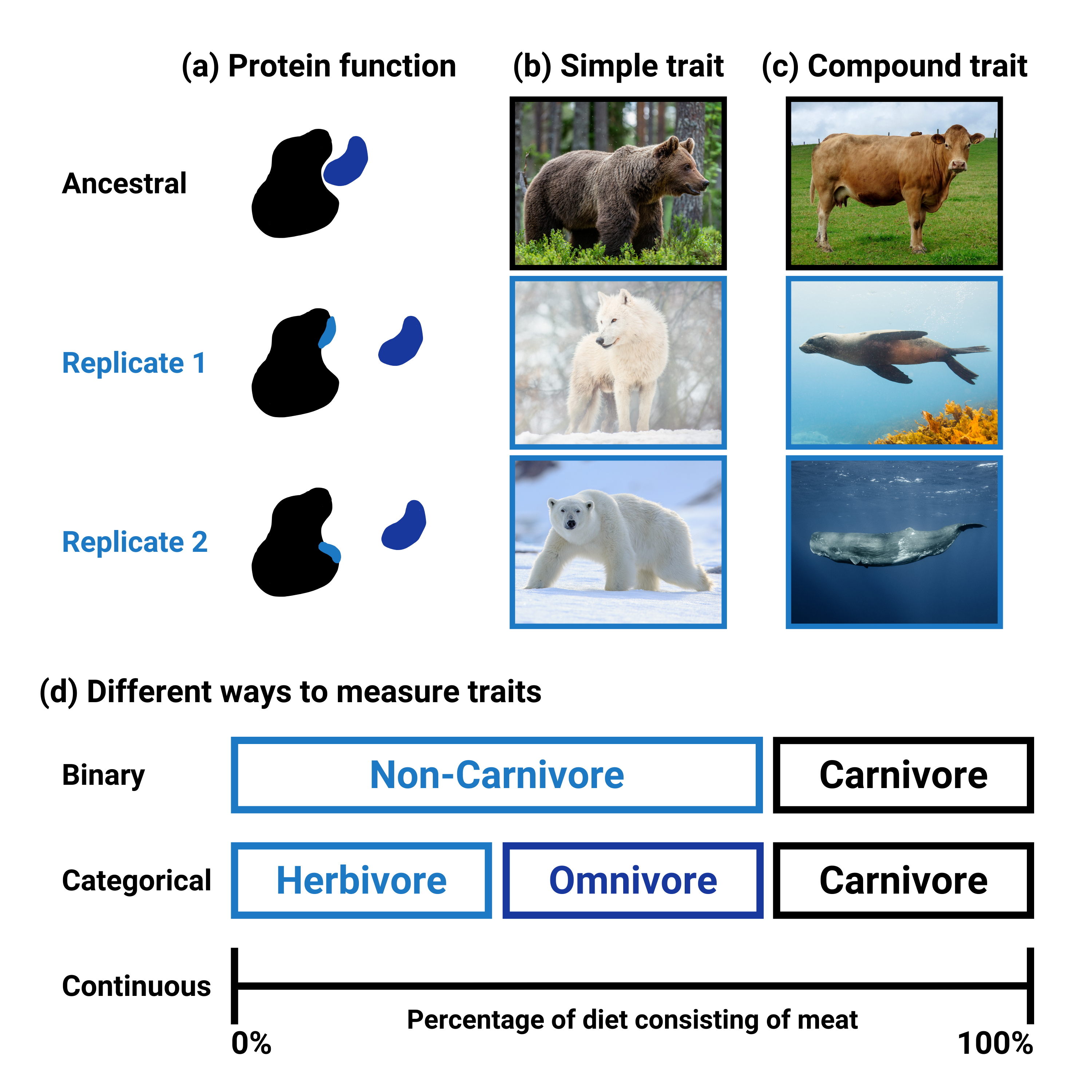}
    \caption{Phenotypic replication can occur at different levels of complexity, and the traits involved can be measured in different ways. Panels (a-c): Three increasing levels of complexity in replicated traits. The top row represents the ancestral phenotype. The bottom two rows are lineages which have independently evolved similar phenotypes. (a) Replicated changes in a simple molecular function. Two different changes in a protein both prevent it from binding to a particular molecule. (b) Replicated changes in a simple macroscopic trait. Animals that have independently transitioned to snowy environments have repeatedly evolved white colouration, such as arctic wolves and polar bears. (c) Changes in multiple simple traits that can be treated as replicated changes in a compound trait. Independent mammalian lineages, e.g. seals and whales, have independently transitioned to living in marine environments.  Panel (d): Traits can be measured using different numbers of categories, or a continuous value. Images in panels (a), (b) and (c) adapted with permission from Shutterstock and Adobe.}
    \label{fig:phenotypic-replication}
\end{figure}

The replicated evolution of a phenotype can occur across varying levels of trait complexity (Figure \ref{fig:phenotypic-replication}) \citep{Speed2016}. In a simple case, the function of a protein may change in similar ways in independent lineages in response to a common environmental stressor. For example, several insect species have independently evolved resistance to a toxic compound produced by their shared host plant \citep{Zhen2012}. Some of these insect species evolved similar or identical changes in a protein which were shown to provide resistance to the toxin (Figure \ref{fig:phenotypic-replication} a). In a slightly more complex example, replication may be observed in changes to the size or colour of specific body parts. This is seen, for example, in the replicated evolution of white colouration in arctic animals \citep{Losos2011} (Figure \ref{fig:phenotypic-replication} b). In even more complex cases, replication can occur across many traits simultaneously when species transition to novel environments. Examples of this level of phenotypic replication include the independent evolution of C4 and CAM photosynthesis in plants that have transitioned to warm and dry environments \citep{Heyduk2019} and mammals transitioning to marine environments \citep{Foote2015} (Figure \ref{fig:phenotypic-replication} c). 

Identifying whether species have undergone replicated evolution depends on how their relevant traits are measured \citep{Grossnickle2024, Redlich2024}. Replication is often identified in a presence/absence (or retained/lost) binary phenotype. However, these categorisations can omit the nuance underpinning traits of interest, limiting the insights gained \citep{Lamichhaney2019}. The value of including more phenotypic categories has been demonstrated by \citet{Redlich2024}. In their analysis of mammalian diets they found that including three categories (herbivore, omnivore, carnivore) rather than just (carnivore, non-carnivore) increased the power of the RERconverge method to identify genetic changes associated with diet (Figure \ref{fig:phenotypic-replication} d). Beyond an increase in the number of categories, traits can also be measured as continuous. For such traits, replication can be defined as distinct lineages independently undergoing either convergence to the same trait value, or a similar direction of trait change \citep{Baker2021, Gemmell2024}. Treating continuous traits as continuous rather than lumping them into categories seems likely to have better statistical power (Figure \ref{fig:phenotypic-replication} d).

Complex traits that are treated as having binary phenotypes are frequently composed of a number of simpler traits that may be categorical or continuous in nature \citep{Lamichhaney2019}. Species that have the same binary value of the compound trait may have distinct values of the underlying simple traits. This makes it challenging to identify the genetic elements that contribute to the phenotype in the compound trait. An example of a system that could benefit from careful consideration of the traits involved is that of marine-adapted mammals \citep{Foote2015}. This adaptation is typically defined to consist of three lineages, namely \textit{Cetacea} (whales and dolphins), \textit{Pinnipedia} (seals and walruses), and \textit{Sirenia} (dugongs and manatees), which have all independently transitioned from terrestrial ancestors to a predominantly or entirely marine lifestyle. Many studies have attempted to identify genetic changes associated with this phenotypic transition (e.g. \citet{Zhou2015, Chikina2016, Hu2019} and \citet{Treaster2021}). However, the compound trait of ``marine adapted" is incredibly complex, consisting of a large number of simpler traits. Many of these simple traits are not shared between all three lineages and might be shared with species that are not ``marine adapted". Because of this, some of the genetic changes associated with the compound trait ``marine adapted'' may not be identifiable by exclusively studying marine lineages \citep{Speed2016, Lamichhaney2019}. Instead, identification of these genetic changes may require multiple studies of individual traits taking into account all mammalian lineages where they undergo replicated changes (e.g. \citet{Wang2024}).

\subsection{Replicated Evolution of Genetic Elements}

\begin{figure} 
    \centering
    \includegraphics[width=1\linewidth]{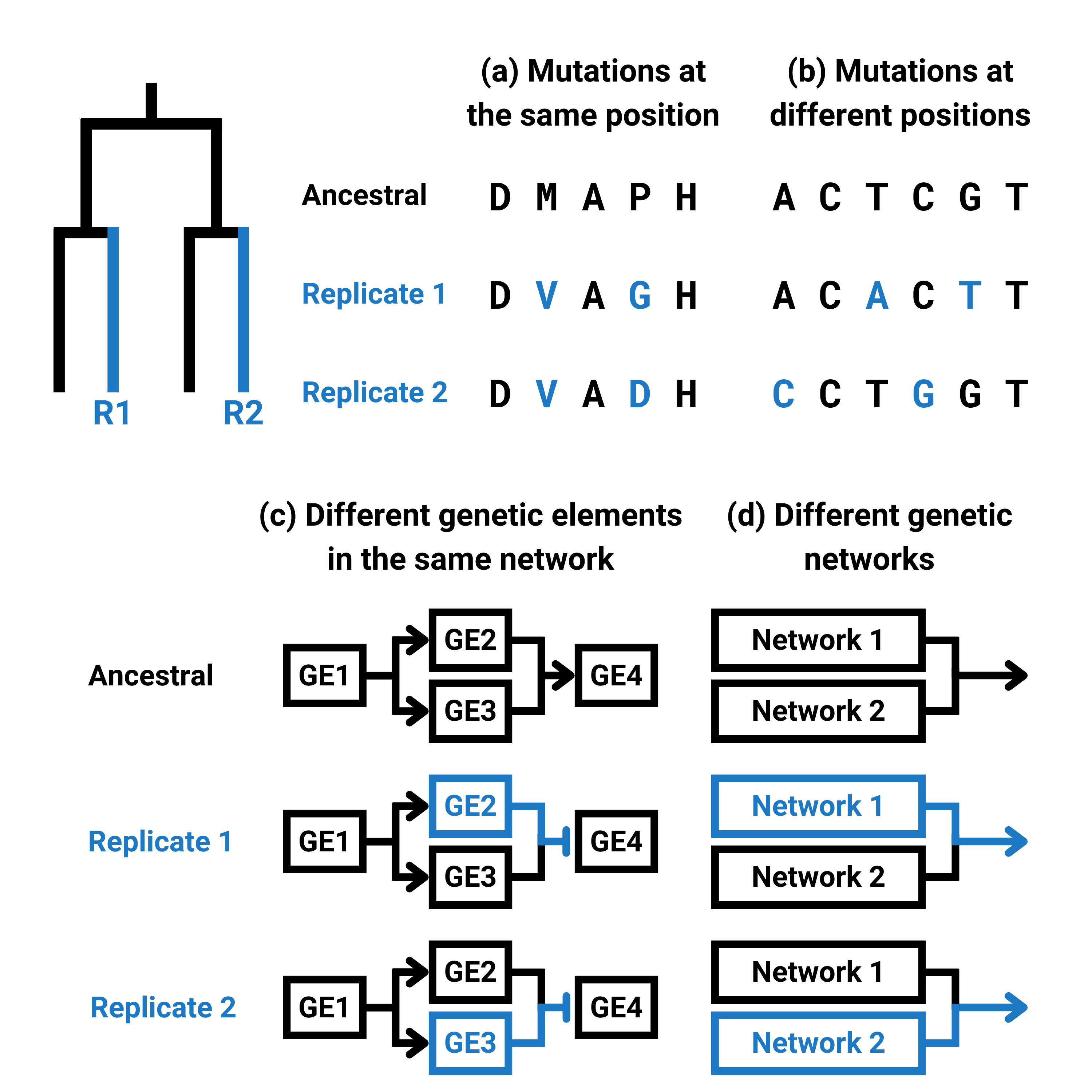}
    \caption{Genetic changes that produce replicated changes in phenotype can occur at various levels of genetic organisation. (a) Change of the amino acid at a site can cause replicated changes in phenotype. (b) When multiple mutations have the capacity to create the change in phenotype, then different mutations within the same genetic element (e.g. in a single gene) can produce replicated changes in phenotype. (c) Similarly, multiple mutations in different genetic elements within the same genetic network may be able to produce a replicated change in phenotype. (d) Some traits may be controlled by multiple networks, and replicated changes in phenotype could be driven by mutations in different networks.}
    \label{fig:genetic-replication}
\end{figure}

Phenotypic replication in independent lineages can be caused by changes that occur at different genomic scales \citep{Losos2011}. For instance, identical mutations at the same nucleotide position may cause identical changes in phenotype (Figure \ref{fig:genetic-replication} a), such as toxin resistance in insects \citep{Zhen2012}. Changes in different sites within the same genetic element may also produce a replicated phenotype (e.g, a gene or regulatory element) (Figure \ref{fig:genetic-replication} b). This could occur when there are multiple ways to achieve a similar change in protein structure \citep{Sykes2022}. Beyond the level of a single genetic element, mutations in different genetic elements within the same genetic network may also lead to a replicated change in phenotype if the different mutations produce similar changes in the overall network function \citep{Laruson2020} (Figure \ref{fig:genetic-replication} c). Finally, genetic replication may not be observed at all when mutations in different networks, and potentially unrelated regions of the genome, are able to create similar changes in a phenotype (Figure \ref{fig:genetic-replication} d). Similarly, we would expect low power to detect genetic replication for traits controlled by a very large number of loci of small effect (i.e. the infinitesimal model).

The degree of pleiotropy \citep{Zhang2023} (where a single genetic element can affect multiple traits) and genetic redundancy \citep{Laruson2020} (where multiple different mutations can have similar impacts on a trait) in genetic elements and the networks they form can also vary substantially. This is important to consider when investigating the genetic causes of phenotypic replication, as the likelihood of observing corresponding genetic replication depends on the nature of the genetic networks controlling the phenotype \citep{Stern2013}. For traits controlled by genetic elements with a high level of pleiotropic constraint, mutations are likely to have substantial deleterious impacts on other traits. Hence, there may be very few viable mutations that are able to change the trait of interest without  disrupting the functions of other traits, giving a higher chance of observing replicated genetic changes. In these cases, genetic replication may be more likely in the regulatory regions controlling the genes in the network. This is because regulatory regions are less likely to be pleiotropic, therefore changes in them are less likely to deleteriously impact other traits controlled by their target gene \citep{Zhang2023}. Conversely, a trait may be controlled by many genetic elements with a high level of redundancy and low pleiotropic constraint. This is especially likely following a duplication of a genetic element, as a particular function of the genetic element can be maintained in one copy even if it is disrupted by a mutation in the other. In these cases, there may be a large number of mutations that could cause a given change in a trait, and it may be surprising to see any replication at the genetic level \citep{Zou2015}. 

\begin{figure} 
    \centering
    \includegraphics[width=1\linewidth]{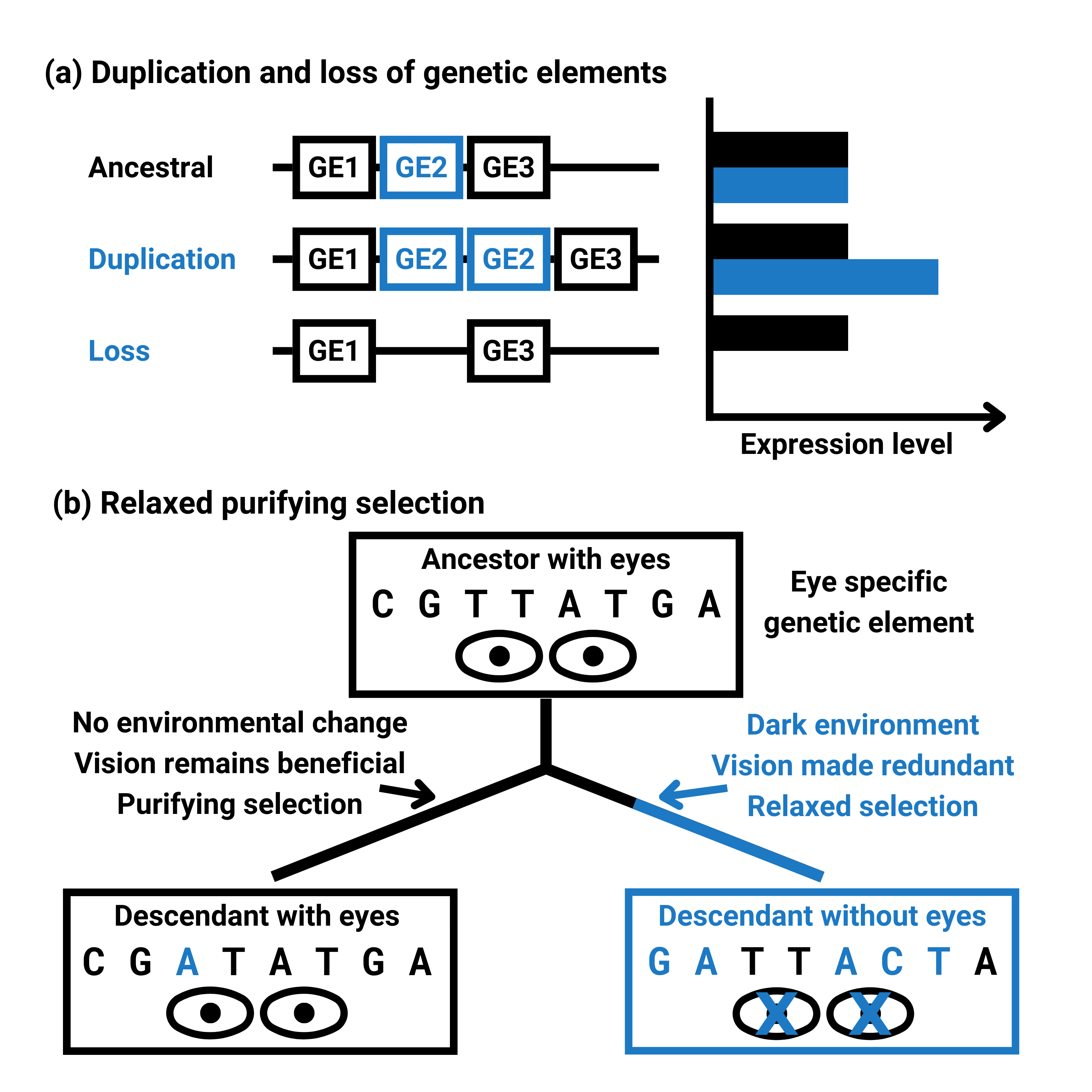}
    \caption{(a) Changes in the number of copies of a genetic element can cause changes in gene expression, which in turn can produce changes in phenotype. Replicated increases and decreases in the number of copies of a gene may produce similar phenotypic changes. (b) When a trait becomes redundant, such as vision in a dark environment, the selective pressure on non-pleiotropic genetic elements that control the trait will relax. This causes mutations to accumulate, leading to the eventual loss of the trait. In contrast, purifying selection will continue to act on genetic elements in lineages where the trait has not become redundant. When the same trait becomes redundant in multiple independent lineages, replicated divergence of the associated genetic elements and subsequent trait loss may occur.}
    \label{fig:duplication-and-loss}
\end{figure}

In addition to the role that duplication of genetic elements can play in reducing pleiotropic constraint, a change in the number of copies of a genetic element can itself create functional changes in phenotype (Figure \ref{fig:duplication-and-loss} a) \citep{Nagy2014, Nagy2016}. For example, more copies of a genetic element can lead to increased expression levels, which in turn can produce a substantial change in a trait. Additionally, if multiple genetic elements play a similar role in controlling a trait, then a particular change in that trait could be caused by duplication of any one of those genetic elements \citep{Griesmann2018}. As such, the likelihood of observing replication in the genetic elements which are duplicated is again dependent on pleiotropy and redundancy. The increased expression of duplicated genetic elements can have deleterious impacts on other traits, and the extent to which this will occur depends on how the regulatory processes within the network respond to the change. Where duplication has driven phenotypic change, replicated reduction in the number of copies of the genetic element and subsequent loss of the changed trait value can occur \citep{Nagy2016, Griesmann2018}.

The replicated loss of entire traits is of particular interest in the context of genotype to phenotype mapping. When the environment occupied by a lineage changes, certain traits may become redundant (Figure \ref{fig:duplication-and-loss} b). As a result, the selective pressure maintaining the function of the genetic elements which control those traits may relax to the point that they evolve under neutral selection, or for elements that are pleiotropically linked to other traits they may evolve under relaxed purifying selection \citep{Hiller2012}. When these losses occur independently in multiple lineages, they may all start to show elevated rates of substitution in those genetic elements relative to other lineages for which the trait has been maintained. This type of replication has proven to be very useful for genotype to phenotype mapping using PhyloG2P methods \citep{Prudent2016, Sackton2019}. However, identifying a replicated loss in phenotype does not guarantee that corresponding degradation of genetic elements will be observed \citep{Womack2023}. This is because traits can be lost over short evolutionary time spans following changes in development. All traits of an organism are formed during development, and small changes in developmental timing can lead to the trait not forming \citep{Womack2018, Pereira2023}. In cases where traits have been lost due to developmental changes, the change in selective pressure on associated genetic elements will still occur, but their impact may not be noticeable for many generations following the initial loss. As such, relevance of trait loss to genotype to phenotype mapping is dependent on sufficient time having passed such that the impacts of relaxed selective pressure are detectable.

There are many phenotypic and genetic levels at which replicated evolution can occur. We have discussed the reasons why genetic replication may be more difficult to identify for replication in compound traits. However, there is still no clear consensus on which forms of genotypic replication are most likely to drive observed phenotypic replication at various scales. Indeed, there have been examples of closely related lineages with replication in simple traits that show little replication of genetic elements \citep{James2021a}, and distantly related species with replication in compound traits that have been shown to have genetic replication at the level of single amino acid substitutions \citep{Liu2014}. Gaining a clearer understanding of how genotypic and phenotypic replication are related will improve our understanding of the genotype to phenotype map, as well as deepen our insights into adaptation and evolution more broadly \citep{Stern2013}. The PhyloG2P research program is well placed to provide novel insights into this relationship. 

\subsection{How population genetics quantifies the genetic basis of replicated evolution}

Before delving into the specifics of PhyloG2P, we first briefly cover population genetics approaches that have been employed to understand the relationship between phenotypic and genotypic replication, as well as genotype to phenotype mapping more broadly.

Population genetics approaches have been widely used to identify regions of the genome associated with repeated adaptation to similar selective pressures. By examining how allele frequencies change overtime within and between populations, population genetics provides insights into how evolutionary forces (such as natural selection, genetic drift, and gene flow) shape the genomes of individuals. As with PhyloG2P approaches, population genetics can help uncover the genetic basis of adaptation at different levels of biological organisation (Figure \ref{fig:genetic-replication}) including nucleotide sites, individual genes, functional pathways, and gene networks (e.g. \citet{James2021a, Rivas2018, Bohutinska2024}). Such approaches range from genome scans in natural populations \citet{Cruickshank2014, Ravinet2017} to genetic mapping applied to experimentally constructed populations, including Quantitative Trait Loci (QTL) mapping \citep{Doerge2002} and Genome Wide Association Studies (GWAS) \citep{Uffelmann2021}. Although these approaches are advancing our understanding of the relationship between the genotype and phenotype, they focus on variation between individuals within populations as well as between closely related species. As a result, population genetics studies often lack the capacity for broad-scale inferences across distantly related lineages, limiting insights into how traits and genes are associated across deeper phylogenetic scales \citep{Lamichhaney2019, Nagy2020, Smith2020}.

\section{An Overview of Current PhyloG2P Methods}

PhyloG2P approaches expand on the genotype to phenotype mapping capabilities of population genetics by examining phenotypic variation across distantly related lineages. While phylogenetic methods that identify correlations between genetic and phenotypic changes have existed for several decades \citep{Zhang1997}, recent advancements in PhyloG2P approaches are characterised by two key features. The first is the ability to analyse sequence data on a genomic scale. The recent genomics gold rush is producing an abundance of genome-wide sequencing data for large numbers of species. This scale of genomic data enables researchers to discover novel genotype to phenotype relationships, without prior functional knowledge of the genetic elements \citep{Nagy2020}. The second feature is the utilisation of replicated evolution, as discussed above. However, many existing methods to identify patterns of selection and adaptation lack both genome-wide capacity and replication-based approaches, and are not well suited for identifying novel genotype to phenotype maps. We focus our review on methods which can be applied at a genomic scale and take advantage of replication, as they represent the most promising approach for expanding existing genotype to phenotype mapping beyond the scale of population genetics.

Many PhyloG2P methods have their roots in the early work of either \citet{Zhang1997} or \citet{Yang1998a}. \citet{Zhang1997} developed a method for detecting replicated amino acid substitutions (Figure \ref{fig:genetic-replication} a) and determining if the number of replicated changes observed was greater than expected by chance for the given phylogeny. In contrast, the branch model developed by \citet{Yang1998a} (now available as part of PAML \citep{Yang2007}) allows for the identification of phylogenetic branches that have increased rates of evolution for a given coding gene (as measured by the ratio of synonymous to non-synonymous substitutions). Both of these and several other early methods \citep{Castoe2009, Pollard2010, Lartillot2010} have been used extensively in PhyloG2P analyses. However, most of these methods were not originally designed for PhyloG2P application and therefore have some limitations. One challenge lies in the computational resources required to analyse large numbers of genetic elements, especially for genome-wide data involving many species \citep{Kumar2022}. Working on a genomic scale also requires accurate estimation of the likelihood of replicated genetic changes occurring by chance \citep{Zou2015}. Additionally, many of these methods are only applicable to coding genes, and cannot capture changes in non-coding regions which could play an important role in trait changes. These challenges have prompted the development of a range of new PhyloG2P methods designed to address these issues and expand the scope of genotype to phenotype mapping (Table \ref{tab:phylog2p_methods}).

We argue that PhyloG2P methods fall into three broad categories based on the type of genomic changes they identify. The first group search for replication at individual amino acid sites. There are a variety of approaches taken to determine which amino acid changes constitute replication, and when observed replication is significantly greater than expected by chance. The second group of approaches search for replicated changes in the rate of genetic evolution. Decreases in rate can be associated with novel constraints on genetic elements, while increases in rate can be indicative of directional or relaxed selection. The third group search for repeated duplication or loss of genetic elements, encompassing a variety of genetic phenomena and phenotypic transitions associated with these changes. Here, we provide a summary of the current PhyloG2P methods for each of these categories (Table \ref{tab:phylog2p_methods}), building on the earlier reviews of \citet{Smith2020, Nagy2020} and \citet{Pereira2023}. We also include some of the key methods mentioned above that cannot be applied on a genomic scale, but are still useful for characterising genetic elements identified by PhyloG2P methods. 

In our discussion, we use the language of ``foreground" lineages to refer to groups of independent lineages which have repeatedly evolved a particular phenotype of interest and ``background" lineages to refer to all lineages without the replicated phenotype. This language can also be extended to foreground and background branches to refer to internal branches of the phylogeny for which an estimate of the trait value is made.

\textls[0]{
\begin{table}
\resizebox{\textwidth}{!}{
\noindent\begin{tabular}{| m{0.13\linewidth} | m{0.49\linewidth} | m{0.04\linewidth} | m{0.04\linewidth} | m{0.04\linewidth} | m{0.26\linewidth} | }

\multicolumn{6}{l}{\textbf{Individual Amino Acids}} \\ 
\hline
\textbf{Model} & \textbf{Description} & \multicolumn{1}{l|}{\textbf{\begin{tabular}[c]{@{}l@{}}\footnotesize{Genome} \\ \footnotesize{Scale}\end{tabular}}} & \multicolumn{1}{l|}{\textbf{\begin{tabular}[c]{@{}l@{}}\footnotesize{Non-} \\ \footnotesize{Coding} \\ \footnotesize{Regions}\end{tabular}}} & \multicolumn{1}{l|}{\textbf{\begin{tabular}[c]{@{}l@{}}\footnotesize{Relaxed} \\ \footnotesize{Match}\end{tabular}}} & \textbf{\begin{tabular}[c]{@{}l@{}}Example \\ Applications\end{tabular}} \\ 
\hline
Zou2015 & Identifies coding genes with a greater number of replicated amino acid substitutions than would be expected by chance \citep{Zou2015} & \makebox[0pt]{\ding{51}} & \makebox[0pt]{\ding{53}} & \makebox[0pt]{\ding{53}} & \citet{Wang2024}\\ 
\hline
WGCCRR & Web application similar to Zou2015 but with more stringent match conditions \citep{Dong2024} & \makebox[0pt]{\ding{51}} & \makebox[0pt]{\ding{53}} & \makebox[0pt]{\ding{53}} &  \\ \hline
Castoe2009 & Compares the number of replicated substitutions to the number of divergent substitutions between two lineages \citep{Castoe2009} & \makebox[0pt]{\ding{51}} & \makebox[0pt]{\ding{53}} & \makebox[0pt]{\ding{53}} & \begin{tabular}[c]{@{}l@{}}\citet{Thomas2015},\\ \citet{Xu2021}\end{tabular} \\ 
\hline
Chabrol2018 & Determines the likelihood of replicated amino acid substitutions having occurred \citep{Chabrol2018} & \makebox[0pt]{\ding{51}} & \makebox[0pt]{\ding{53}} & \makebox[0pt]{\ding{51}} &  \\ \hline
CONDOR & Web application similar Zou2015, but with the option to only require some foreground lineages to have replicated substitutions \citep{Morel2024} & \makebox[0pt]{\ding{51}} & \makebox[0pt]{\ding{53}} & \makebox[0pt]{\ding{51}} &  \\ 
\hline
PCOC & Identifies amino acid positions where a change in amino acid profile has occurred for foreground lineages \citep{Rey2018} & \makebox[0pt]{\ding{53}} & \makebox[0pt]{\ding{53}} & \makebox[0pt]{\ding{51}} & \begin{tabular}[c]{@{}l@{}}\citet{Eliason2023},\\ \citet{Zhang2023a}\end{tabular} \\ 
\hline
CSUBST & Similar to Zou2015, but compares replicated synonymous and non-synonymous substitutions to determine significance \citep{Fukushima2023} & \makebox[0pt]{\ding{51}} & \makebox[0pt]{\ding{53}} & \makebox[0pt]{\ding{53}} & \begin{tabular}[c]{@{}l@{}}\citet{Sadanandan2023},\\ \citet{Matsuda2024},\\ \citet{Morales2024}\end{tabular} \\ 
\hline

\noalign{\vskip 2mm}
\multicolumn{6}{l}{\textbf{Evolutionary Rates}} \\ 
\hline
\textbf{Model} & \textbf{Description} & \multicolumn{1}{l|}{\textbf{\begin{tabular}[c]{@{}l@{}}\footnotesize{Genome} \\ \footnotesize{Scale}\end{tabular}}} & \multicolumn{1}{l|}{\textbf{\begin{tabular}[c]{@{}l@{}}\footnotesize{Non-} \\ \footnotesize{Coding} \\ \footnotesize{Regions}\end{tabular}}} & \multicolumn{1}{l|}{\textbf{\begin{tabular}[c]{@{}l@{}}\footnotesize{Cont.} \\ \footnotesize{Traits}\end{tabular}}} & \textbf{\begin{tabular}[c]{@{}l@{}}Example \\ Applications\end{tabular}} \\ 
\hline
PAML & Compares ratio of synonymous to   non-synonymous substitutions in foreground and background lineages   \citep{Yang2007} & \makebox[0pt]{\ding{51}} & \makebox[0pt]{\ding{53}} & \makebox[0pt]{\ding{53}} & \begin{tabular}[c]{@{}l@{}}\citet{Eliason2023}, \\ \citet{Zhang2023a},\\ \citet{Matsuda2024}, \\ \citet{Wang2024}\end{tabular} \\ 
\hline
phyloP & Generalised framework for   testing hypotheses about the rate of sequence evolution \citep{Pollard2010} & \makebox[0pt]{\ding{51}} & \makebox[0pt]{\ding{51}} & \makebox[0pt]{\ding{53}} & \begin{tabular}[c]{@{}l@{}}\citet{Feigin2019}, \\ \citet{Sakamoto2024}\end{tabular}\\ 
\hline
\begin{tabular}[c]{@{}l@{}}RELAX, \\ aBSREL,\\ TraitRepProp,\\ TraitRELAX\end{tabular} & Detailed methods for determine   the type of selective pressure a coding gene has evolved under (see text for citations) & \makebox[0pt]{\ding{53}} & \makebox[0pt]{\ding{51}} & \makebox[0pt]{\ding{53}} & \begin{tabular}[c]{@{}l@{}}\citet{Sadanandan2023}, \\ \citet{Zhang2023a},\\ \citet{Wheeler2022},\\ \citet{Eliason2023}\end{tabular} \\ 
\hline
\begin{tabular}[c]{@{}l@{}}Forward\\ Genomics\end{tabular} & Identifies genetic elements with a high level of sequence divergence in lineages that have lost a trait \citep{Prudent2016} & \makebox[0pt]{\ding{51}} & \makebox[0pt]{\ding{51}} & \makebox[0pt]{\ding{53}} &  \\ 
\hline
RERconverge & Identifies genetic elements with   significantly different relative evolutionary rate in lineages with a   replicated trait \citep{Kowalczyk2019} & \makebox[0pt]{\ding{51}} & \makebox[0pt]{\ding{51}} & \makebox[0pt]{\ding{51}} & \begin{tabular}[c]{@{}l@{}} \citet{Eliason2023},\\ \citet{Morales2024},\\ \citet{Zhang2023a}\end{tabular} \\ 
\hline
TRACCER & Similar to RERconverge, but   based on pair-wise comparisons with more closely related pairs weighted   higher \citep{Treaster2021} & \makebox[0pt]{\ding{51}} & \makebox[0pt]{\ding{51}} & \makebox[0pt]{\ding{53}} &  \\ 
\hline
PhyloAcc & Similar principle to PAML, but   using evolutionary rates of non-coding regions \citep{Thomas2024} & \makebox[0pt]{\ding{51}} & \makebox[0pt]{\ding{51}} & \makebox[0pt]{\ding{51}} & \citet{Sackton2019} \\ 
\hline

\noalign{\vskip 2mm}
\multicolumn{6}{l}{\textbf{Duplication and Loss of Genetic Elements}} \\ 
\hline
\textbf{Model} & \textbf{Description} & \multicolumn{1}{l|}{\textbf{\begin{tabular}[c]{@{}l@{}}\footnotesize{Genome} \\ \footnotesize{Scale}\end{tabular}}} & \multicolumn{1}{l|}{\textbf{\begin{tabular}[c]{@{}l@{}}\footnotesize{Non-} \\ \footnotesize{Coding} \\ \footnotesize{Regions}\end{tabular}}} & \multicolumn{1}{l|}{\textbf{\begin{tabular}[c]{@{}l@{}}\footnotesize{Dup.}\\ \footnotesize{and/or} \\ \footnotesize{Loss}\end{tabular}}} & \textbf{\begin{tabular}[c]{@{}l@{}}Example \\ Applications\end{tabular}} \\ 
\hline
COMPARE, CAFE & Identifies gene families that undergo duplications of losses correlated with phenotypic changes \citep{Nagy2014}, \citep{Mendes2020} & \makebox[0pt]{\ding{51}} & \makebox[0pt]{\ding{51}} & \makebox[0pt]{Both} & \begin{tabular}[c]{@{}l@{}}\citet{Nagy2016},\\ \citet{Zhang2023a}\end{tabular} \\ 
\hline
REforge & Uses a model of transcription factor binding to   determine when a binding site has deteriorated in lineages that have lost a   trait \citep{Langer2018} & \makebox[0pt]{\ding{51}} & \makebox[0pt]{\ding{51}} & \makebox[0pt]{Loss} &  \\ 
\hline
Evolink & Associated presence/absence of   genetic elements with presence/absence of traits in microbial lineages \citep{Yang2023} & \makebox[0pt]{\ding{51}} & \makebox[0pt]{\ding{51}} & \makebox[0pt]{Loss} &  \\ 
\hline

\end{tabular}}
\caption{Current PhyloG2P methods grouped into three categories. Genome Scale: method has been designed to run at genome-wide scale. Non-Coding Regions: method can be applied to non-coding regions. Relaxed Match: Match between trait and amino acid does not have to be perfect. Continuous Traits: method can be applied to continuous traits. Duplication and/or Loss: if method can detect both phenomena or just loss. Note, some methods do not have example applications apart from the original paper.}
\label{tab:phylog2p_methods}
\end{table}
}

\subsection{Replicated Amino Acid Substitutions} \label{subsec:phyloG2P_single_site}

PhyloG2P methods based on replication at the level of single amino acids (Figure \ref{fig:genetic-replication} a) search for sites with repeated amino acid substitutions in lineages with a replicated binary phenotype of interest. These methods aim to identify coding genes which show an excess of repeated amino acid substitutions \citep{Zhang1997, Chabrol2018, Fukushima2023, Morel2024}, with each approach differing in how they define replicated changes in amino acids. The earliest statistical method of identifying this form of genetic replication was introduced by \citet{Zhang1997}, and subsequently revised by \citet{Zou2015}. These approaches use reconstructed ancestral sequences to identify sites that have the same amino acid in foreground lineages but differ to the amino acid of the most recent ancestor of each lineage. The significance of observed amino acid replication is based on expected number of replicated substitutions given the amino acids likely to be observed at that site (see \citet{Zou2015} for a revised approach to this process). Recently, Whole Genome Comparative Coding Region Read (WGCCRR, \citet{Dong2024}) has been developed as a web-based tool that can facilitate this type of PhyloG2P analysis. However, the WGCCRR framework has the additional condition of requiring background lineages to have different amino acids to those present in the foreground lineages (discussed further in Section \ref{subsec:genotypic_phyloG2P_considerations}).

\citet{Castoe2009} propose a method with a similar definition of replication to \citet{Zhang1997}, but with a different approach to identify statistical significance. In addition to counting the number of replicated changes of the same amino acid, they also count the number of divergent changes, where the foreground lineages have different amino acids at a particular site. They take the ratio of these values as their test statistic and construct a null distribution by computing it many times for random combinations of lineages. They argue that if the ratio calculated for the foreground lineages is significantly greater than the ratios calculated for other groups of species, then the observed replication in the foreground lineages is significant. While the original study by \citet{Castoe2009} focussed on genome-wide rates of replication, subsequent studies have used the same approach to identify specific coding genes associated with repeated phenotypic evolution \citep{Thomas2015, Xu2021}.

The methods of both \citet{Zou2015} and \citet{Castoe2009} require that the foreground lineages share the same amino acid. However, there are cases where replicated changes in protein function arise from different amino acids at the same site \citep{Zhen2012, Mohammadi2022}. One approach to address this is to relax the requirement for each foreground lineage to have evolved the same amino acid. For example, this can be achieved running the above two methods multiple times for different subsets of the foreground lineages. However this approach is computationally intensive for studies with a large number of foreground lineages \citep{Fukushima2023}. \citet{Chabrol2018} developed a model that inherently accommodates situations where some foreground species have different amino acids. Their approach steps away from explicit ancestral sequence reconstruction and instead uses a Continuous Time Markov Model of sequence evolution to determine the likelihood of replicated amino acid transitions occurring in the phylogeny branches leading to the foreground lineages. The ConDor workflow (\citet{Morel2024}, available as a web-based tool) can also facilitate some foreground lineages not sharing a particular amino acid by allowing the user to specify a minimum number of lineages required to share the same amino acid substitution. Their approach was designed to account for potential uncertainty in phenotypic classification, which can be common in viral studies. The Profile Change with One Change (PCOC) method \citep{Rey2018} takes a different approach to modeling variation in the amino acids of the foreground lineages by identifying sites where the foreground lineages have a different amino acid preference (e.g. a preference for a hydrophobic versus hydrophilic amino-acid) compared to the background lineages. While the other approaches focus solely on the most common amino acid in the foreground lineages, PCOC uses all the amino acids in the foreground lineage to determine changes in amino acid profile. Additionally, both the ``PC" and ``OC" components of the model can be used independently, allowing for analysis of a range of amino acid replication scenarios.

The recently developed CSUBST model \citep{Fukushima2023} offers a novel approach to analyse replicated amino acid substitutions. While its definition of replication is similar to \citet{Zhang1997} and \citet{Castoe2009}, CSUBST uniquely examines replicated codon changes by independently analyzing synonymous and non-synonymous substitutions. For each coding gene in the alignment, the method divides the observed counts by the expected counts of replicated synonymous and non-synonymous substitutions respectively (\(dN_c = O_c^N / E_c^N,\: dS_c = O_c^S / E_c^S\)). In a manner inspired by the \(\omega\) metric of selective pressure \citep{Yang1998a}, the method takes the ratio of the relative excess of non-synonymous and synonymous replicated substitutions, producing the metric \(\omega_c = dN_c/dS_c\). By accounting for  synonymous substitutions, this metric is robust to errors in topology which can confound the other methods described above \citep{Mendes2016, Mendes2019}. Additionally, as the metric accounts for the expected number of both synonymous and non-synonymous replicated substitutions, it is also robust to poor estimations of those expected numbers, which can pose a challenge for several of the other methods \citep{Zou2015}.

\subsection{Replicated Changes in Evolutionary Rates} \label{subsec:phyloG2P_rates}

PhyloG2P methods based on replicated changes in evolutionary rates search for genetic elements which evolve at different rates in foreground and background lineages. These methods are designed to identify genetic elements in which multiple different mutations could cause a similar change in phenotype (Figure \ref{fig:genetic-replication} b), which may not be identified by methods based on replicated changes at single sites \citep{Treaster2021}. These models vary in how they measure evolutionary rate, what types of rate changes they identify, and which phylogenetic branches they consider for these changes. One of the earliest implementations of this type of method in a PhyloG2P context was the branch model of \citet{Yang1998a} (available in PAML, \citet{Yang2007}). This model works by analysing variation in the ratio of non-synonymous to synonymous substitutions in coding sequences (\(\omega\)) between foreground and background branches of the phylogeny. As synonymous substitutions are expected to arise in a selectively neutral manner, an \(\omega\) value of 1 indicates neutral evolution of the coding gene in a particular branch. \(\omega < 1\) indicates that the coding gene is under purifying selection, while \(\omega > 1\) is a sign of potential positive selection. A likelihood test can compare whether a model with distinct \(\omega\) values for foreground and background branches fits the genetic data better than a single \(\omega\) value for the whole phylogeny. Coding genes which have a higher likelihood under the two rates model likely experienced a change in selective regime following or during the replicated phenotypic transitions, and hence may be associated with the trait. While it was not designed with replicated evolution in mind, the phyloP program \citep{Pollard2010} extends the basic principles of the likelihood test to general substitution rates. Using substitution rates instead of \(\omega\) allows for the analysis of non-coding regulatory elements, although it lacks information on the selective pressure acting on coding genes.

There are a number of other models which utilise \(\omega\) as a measure of selective pressure. While these have been primarily designed to improve analysis of selective and adaptive pressures, they can still be used in a PhyloG2P context, especially for more detailed characterisation of candidate coding genes identified by other PhyloG2P methods. One of the motivators for these methods is that not all codons in a coding gene will be under the same level of constraint, and hence will likely have different values of \(\omega\). Branch-site models (\citet{Zhang2005}, also available in PAML) account for this by allowing each codon to have its own branch model. However, the branch and branch-site models impose specific permissible \(\omega\) values for the background and foreground lineages. To address this, both the RELAX \citep{Wertheim2014} and adaptive branch-site random effects likelihood (aBSREL, \citet{Smith2015}) models assign each branch a discrete distribution of \(\omega\) parameters and each codon can evolve under one of their rates. RELAX allows three \(\omega\) parameters per branch, centered around a value of \(\omega = 1\). A narrow distribution around \(\omega = 1\) indicates that most sites are evolving under approximately neutral selection, while a broad distribution indicates both constraint and directional selection are present. aBSREL assigns a variable number of \(\omega\) parameters to each branch dependent on the branch length, and places no constraint on the \(\omega\) values taken. Replicated changes in the \(\omega\) distributions of either RELAX or aBSREL can then be associated with replicated phenotypic transitions, with the particular distributions providing more detail as to the selective pressures acting on the coding genes. 

Another motivation for more complex models of selective pressure is that trait changes do not necessarily occur at the point of speciation. The models discussed above all implicitly assume this as they assign whole branches as having a particular trait, which can cause erroneous inference of the selective pressures acting on a genetic element \citep{Halabi2020}. One approach to avoid this issue is to jointly model both trait and sequence evolution. In this framework, a continuous time Markov Model of trait evolution is used to infer when changes in phenotype may have occurred (potentially part way along a phylogenetic branch), with the evolutionary rate dependent on the trait value. This was implemented by \citet{OConnor2009} for binary traits, and in CoEvol \citep{Lartillot2010} for continuous traits. This approach was combined with the consideration for between-site variation discussed above in TraitRepProp \citep{LevyKarin2017}, and further extended to account for uncertainty in phenotypic reconstructions in TraitRELAX \citep{Halabi2020}. Jointly modeling trait and sequence evolution requires substantial computation, which has largely prevented these models from being applied in a genome-wide context.
Another potential limitation is that methods of ancestral state reconstruction are likely to be biased in the case where selection is directional (\citet{Holland2020}, discussed further in \ref{subsec:genotypic_phyloG2P_considerations}).

The first rates-based method to be explicitly developed for PhyloG2P applications was Forward Genomics \citep{Hiller2012, Prudent2016}. Forward Genomics specifically identifies genetic elements which have evolved neutrally following replicated loss of a trait (Figure \ref{fig:duplication-and-loss} b). The method works by reconstructing ancestral sequences and searching for genetic elements where foreground lineages have accumulated a relatively large number of changes from the ancestral sequence. The method was revised to account for the background evolutionary rate of lineages, and to apply a more rigorous test of statistical significance \citep{Prudent2016}. This approach was flipped in the ``Reverse Genomics'' approach of \citet{Marcovitz2016}, which searches for genetic elements that had repeatedly accumulated large numbers of substitutions in independent lineages. A trait database is then used to search for traits with a presence-absence pattern that matched the pattern of the presence/absence of divergence for a given genetic element. 

RERconverge \citep{Chikina2016, Kowalczyk2019} is a widely used method that was also explicitly developed for PhyloG2P applications. RERconverge works by constructing a model of the evolutionary rate of each branch accounting for a range of statistical and evolutionary properties \citep{Partha2019}. This model is then compared to the branch specific evolutionary rates inferred from the alignment for each genetic element, with the difference in value being termed the relative evolutionary rate (RER). Ancestral state reconstruction is used to determine the foreground branches with the trait, and the RERs of the foreground branches are compared to the RERs of the background branches for each genetic element. If there is a strong correlation, with the foreground branch RERs being either greater than or less than the background RERs, then the genetic element is inferred to be associated with the trait. The statistical significance of this correlation is determined through extensive simulations of permuted phylogenies (called ``permulations") which aim to capture the distribution of RERs expected to be observed in the context of replicated evolution \citep{Saputra2021}. This approach can be applied to both coding and non-coding genetic elements \citep{Kowalczyk2022}, as well as continuous \citep{Kowalczyk2020} and categorical traits \citep{Redlich2024} (discussed further in Section \ref{subsec:phenotypic_phyloG2P_considerations}). 

While RERconverge has been applied widely \citep{Eliason2023, Morales2024, Zhang2023a}, \citet{Treaster2021} identified two potential issues with the framework which they attempt to address with their model TRACCER. The first issue is that RERconverge requires branches to be assigned as foreground or background through ancestral state reconstruction, which can be challenging for many traits (see Section \ref{subsec:genotypic_phyloG2P_considerations}). TRACCER instead compares extant lineages with and without the phenotype in a pair-wise manner from their most recent common ancestor, removing the requirement to determine which sections of their evolutionary history did and did not have the replicated phenotype. The second issue is that under the RERconverge framework (and many other rates-based methods) the RERs of all background and foreground lineages are weighted equally when determining correlations. \citet{Treaster2021} argue that closely related pairs of foreground and background lineages will have less sequence divergence than distantly related ones, and hence the differences observed between them are more likely to be associated with replicated trait. They further argue that if a background clade that is not closely related to any foreground lineages is highly sampled, then providing equal weighting to the RERs of that clade can mislead correlations of RER values. To address this, TRACCER weights RER significance inversely to the branch length separating the two lineages being compared. Even with these two changes, TRACCER is still likely not robust to the issue of trait change timing discussed above, but it may be less vulnerable to issues with ancestral state reconstruction than RERconverge and some of the other rates-based methods.

Along with RERconverge, PhyloAcc \citep{Hu2019, Thomas2024} is one of the more thoroughly developed frameworks specifically designed for PhyloG2P analysis, with a focus on non-coding genetic elements. PhyloAcc uses a Bayesian approach with three nested models to detect replicated acceleration of genetic evolution in foreground lineages. The simplest model (M0) assumes that all branches of the phylogeny evolve under neutral or constrained evolution, and does not permit any acceleration. This is nested within the M1 model, which allows for acceleration only on the foreground branches. The M1 model is again nested within the M2 model which allows for acceleration on any combination of branches. All three of these models are fit to a genetic element, and the Bayes factors are compared. If M1 is preferred over M0, then that is evidence that acceleration has occurred, and if M1 is preferred over M2, then it is evidence that it has only occurred in foreground lineages. The second comparison with M2 prevents potential false positives resulting from acceleration associated with a non-foreground trait, which are often not accounted for in other rates-based methods. This approach has been expanded to modeling continuous traits \citep{Gemmell2024} and to account for gene tree discordance \citep{Yan2023}, discussed further in Sections \ref{subsec:phenotypic_phyloG2P_considerations} and \ref{subsec:genotypic_phyloG2P_considerations} respectively.

\subsection{Replicated Duplication and Loss of Genetic Elements}

PhyloG2P methods based on replicated duplication and loss of genetic elements search for genetic elements which are repeatedly duplicated or lost in lineages with a replicated phenotype (Figure \ref{fig:duplication-and-loss}). Models such as Computation Analysis of gene Family Evolution (CAFE, \citet{Mendes2020}) reconstruct the phylogenetic history of gene family expansion and contraction. Replicated changes in gene families occurring in foreground branches can indicate that the gene family plays a role in controlling the trait of interest. The comparative phylogenomic analysis of trait evolution (COMPARE, \citet{Nagy2014}) framework is specifically designed to facilitate this type of PhyloG2P analysis. For example, it can be used to identify a gene family which underwent significant expansion in an ancestral lineage that gained a new trait, and subsequently contracted in descendant lineages which have lost the trait (\citep{Nagy2016}). 

The loss of transcription factor binding sites (TFBSs) is a special case of genetic element loss. TFBSs are often contained within larger conserved non-coding regions, and can be difficult to analyse using DNA alignment based methods. If a TFBS acquires several substitutions that change its binding properties in a particular lineage, then a transcription factor that targets it may no longer bind to the regulatory region. However, if substitutions in another section of the regulatory region create a new TFBS that has similar binding properties to the original TFBS, then the transcription factor may still be able to bind to the regulatory region following mutations in the original binding site. Analysis of a DNA alignment of the region would suggest that the TFBS has been lost in this particular lineage, but the properties of the regulatory element have not actually changed. To facilitate analysis of TFBS loss, Regulatory Element forward genomics (REforge \citet{Langer2018}) calculates a binding affinity score between a set of transcription factors and an alignment conserved regulatory element. These scores are also calculated for reconstructed ancestral sequences, providing an estimate for the phylogenetic history of binding affinity. Replicated decreases in binding affinity can then be associated with replicated loss of a trait. REforge was demonstrated to perform better at identifying regulatory elements associated with trait loss than methods based on evolutionary rates. However, it also requires prior knowledge of transcription factors and their binding motifs, which makes it challenging to apply REforge to traits that have not been studied previously.

Finally, we would like to briefly discuss phylogenetic profiling methods \citep{Dembech2023}. These methods are similar to PhyloG2P methods, except that they search for associations between genetic elements rather than between a genetic element and a phenotype. These methods identify genetic elements which are repeatedly lost or conserved together, which indicates that they may have a common function. They are often best suited for analysis of genetic elements associated with fundamental metabolic processes, and can be applied to multiple-kingdom alignments. The basic idea of phylogenetic profiling can be expanded to PhyloG2P analysis by associating the presence and absence of a trait with the presence and absence of genetic elements. This is the idea behind Evolink \citep{Yang2023}, which is specifically designed for PhyloG2P analysis of alignments of tens of thousands of microbial lineages. While the simplistic presence/absence approach would likely fail to detect many of the genetic elements associated with phenotypic change in multi-cellular organisms, it performs well on the simple molecular traits of microbial lineages. The simplified approach also leads to a substantial reduction in computational intensity, which is a barrier to analysis of such large alignments.

\section{Additional considerations for applying PhyloG2P methods}

\subsection{Defining and measuring traits to produce meaningful genotype to phenotype maps} \label{subsec:phenotypic_phyloG2P_considerations}

Phenotypic replication, spanning all levels of trait complexity and phylogenetic scale, has been associated with all scales of genotypic replication \citep{Liu2014, Zhen2012, James2021a, Mohammadi2022, Morales2024}. As the field of PhyloG2P is still young, it is difficult to provide definitive advice on the most effective methods for identifying particular types of genotype to phenotype associations. However, we can discuss the types of phenotypic replication for which PhyloG2P methods are most likely to produce informative genotype to phenotype maps. 

Discovering the genetic changes which underlie the emergence of replicated phenotypes, and hence the genetic elements which control particular traits, is the primary focus of PhyloG2P research \citep{Chikina2016, Morales2024}. For simple traits, the potential discoveries are relatively straightforward. For example, when we ask the question ``Which genetic elements are involved in replicated changes in limb morphology or limb loss?", we seek to discover genetic elements that control particular limb features or are essential to limb development. PhyloG2P studies investigating this question have successfully identified genetic elements that control limbs, and in some cases verified their role through follow-up transgenic experiments \citep{Sackton2019, Booker2016}. In contrast, applying similar questions to compound traits presents additional challenges. For instance, when asking ``Which genetic elements are involved in mammalian transitions to marine environments?", we are aiming to identify genetic elements involved in any one of the many simple traits which have changed during marine transitions \citep{Chikina2016}. In addition to the fundamental issues of identifying relevant genetic elements as discussed in Section \ref{subsec:phenotypic_replication}, the insights gained into the function of any previously uncharacterised genetic elements may be limited. Identified genetic elements could be associated with any of the many basic traits which form the compound ``marine mammal" trait, substantially limiting the specificity of any genotype to phenotype map produced from these studies. Given that PhyloG2P methods have the capacity to create genotype to phenotype maps for genetic elements which have not yet been characterised by population genetics studies, focussing on replication in compound traits may miss opportunities to make meaningful progress on uncovering the ``dark side of the genome" \citep{Nagy2020}. While this issue has been identified previously \citep{Lamichhaney2019}, recent studies are still applying PhyloG2P methods to replication in compound traits, such as flight in vertebrates \citep{Matsuda2024}, echolocation in mammals and birds \citep{Sadanandan2023, Dong2024}, hibernation and vocal learning \citep{Christmas2023}, or broadly defined ecomorphs \citep{CorbettDetig2020, Morales2024}. We strongly encourage research groups considering PhyloG2P analyses of compound traits to instead focus on replication in the component basic traits.

\citet{Lamichhaney2019} suggest that expanding PhyloG2P methods to analyse traits that were non-binary or continuous in nature will assist studies in moving beyond simplistic replication in compound traits. Recent developments have produced PhyloG2P methods capable of analysing continuous valued traits on a genome-wide scale \citep{Kowalczyk2020, Gemmell2024}. RERconverge has introduced an approach to correlate branch RER values with the change in trait value on those branches \citep{Kowalczyk2020}. This framework identifies correlations between any combination of high/low RER scores and increases/decreases in trait values. \citet{Kowalczyk2020} argued that different genetic elements would likely be involved in increases and decreases in trait value, and hence chose to consider both cases individually. A similar directional approach was taken by \citet{TejadaMartinez2021} in their study of longevity in great apes, where they looked for correlation between evolutionary rate and maximum lifespan. In contrast, \citet{Baker2021} searched for correlations between evolutionary rate and both phenotype value and absolute change in trait value. They argued that some genetic elements may be able to drive both increases and decreases in trait value. For example, a high evolutionary rate could cause both large increases and decreases in the trait value for different lineages. A similar approach was taken by \citet{Gemmell2024} to analysing continuous traits in the PhyloAcc framework, where they model changes in continuous trait value as a form of Brownian motion. The rate of this Brownian motion is set by which of the three PhyloAcc rate classes the lineage is evolving under. This will identify similar correlations between evolutionary rate and absolute change in trait value as \citet{Baker2021}, as the increased Brownian rate is more likely to produce large increases and decreases in trait value. Even though the directional and non-directional approaches aim to identify distinct sets of genetic elements associated with the continuous trait, \citet{Gemmell2024} found a considerable overlap between their results and the results of \citet{Kowalczyk2020}. While more studies employing these new methods will be required to fully understand how they may complement each other, they are a promising start to increase the range of traits that PhyloG2P analysis can be applied to. 

Compared to continuous traits, development of PhyloG2P methods for non-binary categorical traits has seen less development. The \(\delta\) statistic of \citep{Ribeiro2023} is one approach that is able to identify associations between genetic elements and variation in a categorical trait. The \(\delta\) statistic measures the uncertainty in ancestral state reconstruction of internal nodes based on a categorical trait for a given phylogeny (such as a gene tree). A low level of uncertainty in ancestral state reconstruction suggests a strong phylogenetic signal, indicating that the genetic element may be associated with the trait of interest. The RERconverge framework has also been expanded to work with categorical traits by using a Continuous Time Markov Model of trait evolution to reconstruct ancestral states \citep{Redlich2024}. This approach conducts pair-wise RER comparisons between all combinations of the traits. The differences in evolutionary rate are assessed using a revised version of their ``permulation" approach to determine statistical significance, and then combined in an omnibus test to identify genetic elements that are correlated with the trait. The \(\delta\) statistic and RERconverge detect different patterns of evolutionary rate change. The \(\delta\) statistic will be large if a genetic element has an increased rate of evolution during transitions between trait values, as the ancestral states of the trait will be inferred with greater certainty. By comparison, RERconverge explicitly identifies genetic elements that evolve at different rates depending on the trait value of the lineage. It has previously been found that these two approaches tend to identify the same genetic elements for a binary trait with few transitions in phenotype \citep{Kowalczyk2022}, but it is not yet clear if the same will be true for non-binary categorical traits with many ancestral transitions in phenotype \citep{Treaster2021}. \citet{Redlich2024} found that \(\delta\) statistic and RERconverge generally identified distinct sets of genetic elements, suggesting that further research into the impacts of these two approaches to modeling changes in evolutionary rate may be required.

In addition to suitable PhyloG2P methods, widespread analysis of basic traits requires easy access to trait data (whether categorical or continuous) \citep{Lamichhaney2019}. While genome-wide alignments are becoming readily available for hundreds of species (e.g. \citet{Christmas2023}), equivalently large databases of detailed trait data remain scarce. The MaTrics database of mammalian trait data has recently been created to help close this gap \citep{Stefen2022}. Designed to represent trait data as categorical values rather than continuous ones, the database facilitates the presence-absence analyses that most PhyloG2P methods have historically focussed on. However, the recent advances discussed above have demonstrated that analysis of continuous traits may play an important role in future PhyloG2P studies. It is likely that these two approaches will compliment each other, as fully uncovering the genetic basis of traits will require analysing both discrete and continuous variation. There are examples of detailed trait databases that contain both categorical and continuous measurements, such as AusTraits \citep{Falster2024}. We recommend that research efforts to collate and digitise existing trait data for the purpose of PhyloG2P analysis capture both categorical and continuous trait value information.

\subsection{Caveats and strengths of different approaches to modeling replication} \label{subsec:genotypic_phyloG2P_considerations}

An issue that has arisen in some PhyloG2P studies is using phylogenetic models that do not accurately represent the type of genetic replication being searched for. One of the most widely identified and discussed occurrences of this is the association of coding genes with the replicated evolution of echolocation in bats and toothed whales by \citet{Parker2013} (see \citep{Thomas2015, Rey2018} for subsequent discussion). \citet{Parker2013} argued that echolocation related genes in bats and toothed whales would likely share more similarities than expected based on the species tree. To attempt to model this, they searched for genes that had higher likelihood values under a phylogeny that grouped the foreground lineages into a clade sister to other mammalian lineages than their likelihood under the species phylogeny. While \citet{Parker2013} intend for this model to represent the biological process of echo-locating lineages independently evolving similar genetic changes, it is instead more likely to represent an alternative hypothesis of the echolocation related genes having a distinct evolutionary history to that of the species tree. It is unclear what biological process this would represent, and as such it is difficult to interpret the evolutionary significance of any genes which are identified with this method. The alternate topology approach also assumes that the entirety of a genetic element is expected to be more similar between foreground lineages, when phenotypic replication may only require substitutions in a subset of sites \citep{LevyKarin2017}. We argue that attempts to identify evolved genetic similarity should use approaches that model this explicitly, such as models of replicated amino acid substitution (see Section \ref{subsec:phyloG2P_single_site}).

The inverse issue to that of \citet{Parker2013} can also occur when incomplete lineage sorting (ILS) or introgression has resulted in apparent replication in phenotypes. When genetic elements have a history of ILS (such as selection on standing variation) or introgression, the phylogeny for the genetic element will differ from the species tree (referred to as gene tree discordance). Forcing such genetic elements onto the species tree can result in apparent increases in substitution rate on branches that differ between the two trees \citep{Mendes2016, Mendes2019}. Evolutionary rates based PhyloG2P models (see Section \ref{subsec:phyloG2P_rates}) can be used to identify genetic elements under these circumstances. However, this would again make it challenging to interpret the evolutionary significance of any genetic elements identified, as these models assume that changes in the genetic element have arisen independently when they have not. When it is possible that ILS or introgression have been involved in phenotypic replication, it may be more appropriate to model those phenomena explicitly. A recent development in PhyloAcc, named PhyloAcc-GT \citep{Yan2023}, analyzes replicated acceleration in evolutionary rates while accounting for the impacts of potential gene tree discordance using a multi-species coalescent model, allowing these two phenomena to be disentangled.

There has been much discussion as to the nature of replicated amino acid substitutions and how they should be modeled in the context of replicated phenotypic evolution \citep{Thomas2015, Zhou2015, Zou2015, Thomas2017, Rey2018, Morel2024}. Central to this discussion is the notion of necessity and sufficiency of particular amino acids underling the replicated phenotype. Necessity describes how a phenotype cannot arise without a specific amino acid substitution, requiring all foreground lineages to share this substitution. On the other hand, sufficiency indicates that any lineage with the amino acid will have the phenotype and hence no background lineages should have this substitution. While these assumptions may hold in specific contexts (such as small viral genomes under strong constraint, \citet{Morel2024}), we argue that they are too restrictive to apply to most biological contexts. Necessity does not hold in cases where multiple amino acids in the same site or substitutions in different sites produce a similar change in phenotype \citep{Rey2018, Sykes2022, Treaster2021}. This restrictive assumption also undermines the replicated/divergent distinction of \citet{Castoe2009} and \citet{Thomas2015}, as they assume that a replicated trait change cannot be caused by two different amino acids in the same position. Likewise, sufficiency does not hold when multiple substitutions are required to produce the replicated phenotype, or where the phenotypic effect of particular amino acid substitutions depends on the genetic and environmental context. Additionally, these issues are further exacerbated for compound traits \citep{Zhou2015, Lamichhaney2019} or when large numbers of lineages are included in the analysis \citep{Thomas2017}. Care should be taken when applying methods with these types of strict assumptions (such as \citet{Castoe2009} or \citet{Dong2024}), and we instead recommend the use of methods that do not rely on them (such as \citet{Chabrol2018}, \citet{Rey2018} or \citet{Morel2024}).

Another major caveat that applies to most, if not all, PhyloG2P methods is that they rely, explicitly or implicitly, on accurate ancestral state reconstruction of the trait of interest to differentiate foreground and background lineages. However, most reconstruction methods assume an absence of selective pressures, which can lead to inaccuracies  when this does not hold \citep{Holland2020}. For example, if a trait has undergone replicated changes in many lineages then the ancestor of those lineages may be incorrectly inferred to have the replicated trait \citep{Sackton2019, Thomas2024}.

The challenges associated with ancestral state reconstruction do not impact all PhyloG2P methods equally. For instance, ambiguity about the ancestral state is unlikely for replicated degradation or loss of a complex trait when applying methods such as Forward Genomics.  In such a case, once a lineage accumulates multiple degrading substitutions in a genetic element that underlies the lost trait (as modeled by Forward Genomics \citep{Hiller2012}), there is very little chance of those substitutions being reversed so as to re-evolve the trait. Loss of complex traits is likely to follow Dollo's law of irreversibility, where we can be confident that the most recent common ancestor of lineages with and without the trait possessed the trait. Conversely, it is challenging to reconstruct ancestral states of replicated transitions between various states of a complex trait (such as ecomorphs) \citep{CorbettDetig2020, Morales2024}. Without reliable estimation of ancestral states, it is unclear where particular phenotypic transitions (and hence genotypic changes) occurred on the phylogeny. Some studies have attempted to avoid this issue by focusing on the branches leading to extant lineages \citep{CorbettDetig2020, Morales2024}. However, this approach risks overlooking genotypic changes that occurred on internal branches of the tree, potentially creating false negatives which would significantly reduce the power of the analysis \citep{Chabrol2018, Halabi2020, Treaster2021}. While it may not entirely resolve the issue, individually analysing each of the basic traits that constitute the compound trait may improve the capacity to infer ancestral states. Additionally, incorporating fossil records \citep{Lamichhaney2019} or methods that do not heavily rely on ancestral state reconstruction \citep{Treaster2021} may also help. However, uncertainty in ancestral state reconstruction still represents an important limitation, and it's potential impacts should be considered carefully when drawing conclusions from PhyloG2P analysis.

\section{Future Directions}

Phenotypic replication can be caused by replicated changes at several genetic levels in both coding and non-coding genetic elements. As we have discussed, there are a range of PhyloG2P methods available to identify replication on these different genetic levels. The diversity of methods available means that there is no one single approach to determine the replicated genetic changes underlying phenotypic replication. This diversity has been actively embraced with several recent studies applying multiple complementary PhyloG2P methods \citep{Zhang2023a, Eliason2023, Matsuda2024}. However, there have been other recent studies that only use a single PhyloG2P method \citep{Xu2021, Sakamoto2024}. Such studies limit the scope of the conclusions they can draw, and risk missing important genetic elements associated with the trait \citep{Wheeler2022}. Similarly, PhyloG2P studies have revealed the presence of genetic replication in non-coding regions when it was not found in coding genes \citep{Feigin2019, Langer2018, Thomas2024}. As such, we strongly encourage future PhyloG2P studies to use a variety of methods and analyse both coding and non-coding genetic elements whenever possible, and provide clear justifications for their methodological choices. 

There are potential barriers that can prevent the widespread application of PhyloG2P methods. One is the accessibility of PhyloG2P methods from a technical perspective. Models which require many complicated steps may prove inaccessible for researchers without the required background knowledge and skills, limiting their capacity to include it in their analysis. To avoid this, researchers creating PhyloG2P methods should include detailed walk-throughs and user-guides (such as \citet{Kowalczyk2019} and \citet{Thomas2024}). Another barrier is computational intensity. While high computational demands are common in genomic analysis, they can limit the capacity for researchers to analyse genomic data. This limitation not only restricts researchers' ability to apply multiple PhyloG2P methods, it also inhibits the capacity for reproduction and verification of previous findings, eroding an important component of rigorous scientific research \citep{Kumar2022}. Research groups developing PhyloG2P methods should explore ways to reduce the computational intensity of their methods (e.g. \citep{Yan2023} found that tuning model complexity on a gene by gene basis did not impact performance). Computational accessibility can also be provided through publicly available web applications \citep{Ribeiro2023, Dong2024, Morel2024}. Ensuring that PhyloG2P methods are accessible will enhance the capacity for researchers to apply comprehensive sets of methods in their analyses, improving the capacity for PhyloG2P research to uncover novel genotype to phenotype maps.

The PhyloG2P methods discussed in this review either focus on changes in molecular sequences, or gene duplication and loss. However, there are many other epigenetic and transcriptomic features of genomes which can impact phenotypes. These include DNA methylation, chromatin structure, structural rearrangements such as chromosomal inversions, and the interactions between RNA, proteins, and their environment. Including variation in these features can improve phenotype to genotype maps \citep{Ritchie2015}, and incorporating them into PhyloG2P analysis may prove fruitful. One way to approach this is through machine learning models, such as the Evolutionary Sparse Learning (ESL) model of \citet{Kumar2021}. ESL uses binary encoding to enable joint analysis of any genetic feature, such as SNPs or presence/absence of a genetic element, as well as epigenetic or environmental variables. Machine learning processes are then used to narrow down the features to those which are most important in explaining a particular phylogenetic hypothesis, such as the presence of a particular phenotype. As well as its capacity to integrate a variety of variables relevant to phenotypic prediction, ESL is also orders of magnitude less computationally intensive than the PhyloG2P models discussed in this review. However, one shortcoming of using machine learning is that it does not explicitly model evolutionary processes, complicating the interpretation of the features it identifies. This means that it may be most appropriate to combine applications of ESL with other methods that use explicit models of evolution, such as PAML \citep{Yang2007} or PhyloAcc \citep{Hu2019}. This approach may prove beneficial, as the more computationally intensive models will only need to be applied to the subset of genetic elements identified by the computationally efficient ESL (provided ESL is not overly inaccurate in its initial identification of candidate genetic elements). As such, the machine learning approach of ESL provides a first step into modeling a broader range of biological features that may impact phenotypes, and also provides a more computationally accessible option for PhyloG2P analysis.

PhyloG2P methods identify a set of genetic elements correlated with a trait of interest, and follow up studies are required to determine the nature of that relationship. Although infeasible for most species, many PhyloG2P studies use existing genetic annotations to verify the relevance of the candidate genetic elements they identify. This can be done through gene ontology enrichment tests, which have often identified significant enrichment in functions related to the trait of interest \citep{Chikina2016, Nagy2016, Prudent2016, Hu2019, Redlich2024}. However, relatively few applications of PhyloG2P methods have been followed up by transgenic or knockout experiments to verify the role of genetic elements identified in their analysis \citep{Booker2016, Sackton2019, Xu2021}. These experiments provide evidence of causal relationships between a genetic element and a trait \citep{Losos2011}, and demonstrate relevance of the genetic elements identified by PhyloG2P methods. More studies such as these will be required in future to build confidence in the capacity for PhyloG2P methods to produce accurate genotype to phenotype maps.

An important limitation of current PhyloG2P methods is that they rely on a single ``representative" trait value and consensus genome for each lineage, which does not account for the phenotypic and genotypic variation within lineages. Trait values can vary widely within lineages, and this variance will likely not be consistent through time or between lineages. This is particularly important for the analysis of traits with continuous values, as allelic variation between individuals may contribute to significantly different values of the trait (such as height or lifespan). Extending PhyloG2P methods to account for within-lineage variance, such as by modeling changes in trait distributions, may improve their accuracy. It is also likely that allelic variation within populations that have undergone replicated phenotypic evolution will show corresponding signs of adaptation in the genetic elements controlling the trait \citep{Barghi2020}. Finding ways to include this information in PhyloG2P analysis may improve the capacity to identify genetic elements associated with trait variation between species. More broadly, a model that unifies both population genetics and PhyloG2P analysis will pave the way for understanding the connection between evolution on short and long timescales. This gap has been partially bridged by the PhyloGWAS approach of \citep{Pease2016}, and may be more comprehensively closed with a unified mathematical framework such as the one proposed by \citet{Schraiber2024}.

In summary, recent years have seen many developments in the PhyloG2P research program. With the growing abundance of genomic sequences, well-resolved phylogenies, and detailed trait databases, PhyloG2P methods are well positioned to expand genotype to phenotype mapping to include variation between species. A detailed understanding of both phenotypic and genotypic replication will reveal a largely untapped wealth of potential applications of these methods, further enabled by the development of PhyloG2P methods that can be applied to continuously valued traits. While some challenges remain for the integration of epigenetic information and  within population trait variation in these studies, recent developments provide promising approaches to addressing them. There are many potential next steps that the PhyloG2P research program could take, and all of them are likely to reveal exciting new opportunities for expanding the mapping of genotype to phenotype.

\section{Acknowledgments}

This work was partly funded by The Australian Research Council Centre of Excellence for Plant Success in Nature and Agriculture (CE200100015).

\bibliographystyle{apalike}
\bibliography{MacdonaldPhyloG2PReview}

\end{document}